\newcommand{\be}{\begin{eqnarray}}
\newcommand{\en}{\end{eqnarray}}

\documentclass[aps,prb,twocolumn,superscriptaddress]{revtex4-1}

\usepackage{amsmath, amsthm, amssymb,subfigure,graphicx,color,dcolumn}

\begin{document}


\title{Giant anharmonicity suppresses superconductivity in AlH$_3$ under pressure}


\author{Bruno Rousseau}
\email{sckrousb@ehu.es}
\affiliation{Donostia International Physics Center (DIPC),Paseo de Manuel Lardizabal,
           	                             20018, Donostia, Basque Country, Spain}
\affiliation{Centro de Fisica de Materiales CSIC-UPV/EHU, 1072 Posta kutxatila, 
					E-20080 Donostia, Basque Country, Spain}

\author{Aitor Bergara}
\email {a.bergara@ehu.es}
\affiliation{Donostia International Physics Center (DIPC),Paseo de Manuel Lardizabal,
           	                             20018, Donostia, Basque Country, Spain}
\affiliation{Centro de Fisica de Materiales CSIC-UPV/EHU, 1072 Posta kutxatila, 
					E-20080 Donostia, Basque Country, Spain}
\affiliation{Materia Kondentsatuaren Fisika Saila, Zientzia eta Teknologia Fakultatea,
Euskal Herriko Unibertsitatea, 644 Postakutxatila,48080 Bilbo, Basque Country, Spain}



\date{\today}

\begin{abstract}
The anharmonic self energy of two zone boundary phonons were computed
to lowest order for AlH$_3$ in the $Pm\bar 3n$ structure at 110 GPa. 
The wavevector and branch index corresponding to these modes are 
situated in a region of phase space providing most of the electron-phonon coupling.
The self energies are found to be very large and the anharmonic contribution
to the linewidth of one of the modes studied could
be distinguished from the electron-phonon linewidth. 
It is found that anharmonicity suppresses the electron-phonon coupling
parameter $\lambda$, providing a possible explanation for the 
disagreement between experiment and previous theoretical studies
of superconductivity in this system. 
\end{abstract}


\maketitle

\section{Introduction}
It has been suggested four decades ago that elemental hydrogen could form
an exceptionally high T$_c$ superconductor under compression\cite{PhysRevLett.21.1748}
(a recent estimate being 242\thinspace K at 450\thinspace GPa\cite{cudazzo:257001}),
and more recently that it could have very exotic properties, such
as being a metallic quantum liquid\cite{babaev:105301}, or forming
protonic Cooper pairs\cite{babaev:105301,babaev_nature}.
However, metallic hydrogen has long been elusive; the dimers 
persist\cite{PhysRevLett.90.035501} 
and hydrogen remains non-metallic up to a pressure of
320\thinspace GPa\cite{Loubeyre_Hydrogen_nature}.

An alternative route to hydrogen superconductivity has been suggested
in the form of hydrides, where the presence of a heavier
element can act to chemically ``pre-compress" hydrogen, compelling it
to reveal its superconducting properties at 
lower pressures\cite{PhysRevLett.92.187002,feng:017006}.
Because of their large hydrogen content, 
superconductivity in the group IV hydrides has been studied 
extensively\cite{stannane,PhysRevLett.102.087005,PhysRevLett.101.107002,
MartinezCanales20062095,PhysRevB.76.064123,Eremets_sih4,metalization_sih4_pnas}.
Some tri-hydrides also received attention 
recently\cite{Ahuja_trends} and it was suggested that the origin of superconductivity
in these systems could be soft phonons in the vicinity of phase transitions.
Within this context, aluminum hydride (AlH$_3$) under pressure has recently
been studied both theoretically\cite{pickard:144114,kim:100102,goncharenko:045504}
and experimentally\cite{goncharenko:045504}. Using random structure searching, 
a particularly interesting phase of symmetry $Pm\bar 3n$ 
 has been found to be energetically favorable above 
$\sim$\thinspace70\thinspace GPa\cite{goncharenko:045504,pickard:144114}. 
This phase contains two formula
units per cell, with the Al ions forming a body-centered-cubic (bcc) structure
and the hydrogen ions forming linear chains on the faces of the cubic cell.
\textit{Ab initio} calculations also suggested that the electron-phonon
coupling parameter should be fairly large in this phase at 
$\sim$\thinspace110\thinspace GPa 
($\lambda\simeq0.74$), predicting a value of T$_c$$\simeq24$\thinspace K,
\cite{goncharenko:045504} in 
agreement with the general idea that compressed hydrides could be good 
superconductors\cite{PhysRevLett.92.187002}. Very interestingly, however,
no superconducting transition was found down to 4\thinspace K\cite{goncharenko:045504};
the reason for the disagreement between theory and experiment is unclear.

In the present work, we show that the phonon modes which provide most of 
the electron-phonon coupling are actually strongly renormalized by anharmonicity,
which should greatly affect the value of the predicted T$_c$.

In sections \ref{section superconductivity} and \ref{section phonon}, 
basic formulas pertaining to phonon mediated superconductivity and 
phonon anharmonicity are reminded, which also serves to fix the notation.
The \textit{ab initio} calculations performed
are described in section \ref{section computations}, and 
the main results pertaining to anharmonicity are presented in section
\ref{section results}.

\section{Superconductivity}
\label{section superconductivity}
The theory of phonon-mediated superconductivity is well understood\cite{AllenSSP}.
A popular approximation to the superconducting transition temperature
is given by the Allen-Dynes modification of the McMillan formula,
\cite{PhysRev.167.331,PhysRevB.12.905}
\be
	k_B \mbox{T}_c &=& \frac{\hbar \omega_{log}}{1.2}
		\exp\Big[-\frac{1.04(1+\lambda)}{\lambda-\mu^*(1+0.62\lambda)}\Big],
\en
where $\mu^*$ is a parameter of order 0.1 which approximately
accounts for the electron-electron repulsion at the Fermi level (which tends
to weaken Cooper pairs, and thus reduce T$_c$), $\omega_{log}$ is
the logarithmic average of the phonon frequencies and $\lambda$ is the
electron-phonon interaction (or electronic mass enhancement) parameter.
This last parameter in turn is obtained from the Eliashberg spectral function
$\alpha^2F$,
\be
	\lambda = \int_0^\infty d\omega \lambda(\omega);\hspace{5mm}
	\lambda(\omega) = 2\frac{\alpha^2 F(\omega)}{\omega}.
\en
The Eliashberg spectral function can be approximately related to the 
phonon linewidths $\gamma_\nu({\bf q})$ by\cite{PhysRevB.6.2577}
\be
	\alpha^2 F(\omega) &\simeq& \frac{1}{2\pi\hbar} \frac{1}{N(\epsilon_F)}
		\frac{1}{N}\sum_{\bf q,\nu} 
		\frac{\gamma_\nu({\bf q})}{\omega_\nu({\bf q})}
			\delta\big(\omega-\omega_\nu({\bf q})\big),
\en
where $N$ is the number of unit cells in the crystal, $N(\epsilon_F)$ is 
the density of states per unit cell at the Fermi energy, $\bf q$
is a wave vector constrained to the first Brillouin zone (1BZ), $\nu$
is a mode label
and $\omega_\nu({\bf q})$ is the frequency of phonon mode $\bf q\nu$.
This implies that $\lambda$ can also be expressed as
\be
	\label{lambda sum}
	\lambda &=& \frac{1}{N}\sum_{\bf q,\nu} \lambda_{{\bf q}\nu },
\en 
with
\be
	\lambda_{{\bf q}\nu } &=&
	\frac{1}{\pi\hbar} \frac{1}{N(\epsilon_F)}
		\frac{\gamma_\nu({\bf q})}{\omega_\nu({\bf q})^2}.
\en

The usual method employed to obtain $\lambda$ from \textit{ab initio}
calculations is to first obtain the band structure of the system,
second to obtain the phonon frequencies within the Born-Oppenheimer
approximation, and third to obtain effective electron-phonon coupling
parameters. The system is then approximately described in terms
of a Fr\"ohlich Hamiltonian,
\be
\lefteqn{\hat{H} =
        \label{frohlich}
        \sum_{n\bf k\sigma} \epsilon_{n\bf k}
        \hat{c}^\dagger_{n\bf k\sigma}\hat{c}_{n\bf k\sigma}
        +\sum_{\bf q\nu} \hbar \omega_\nu({\bf q})
        \Big(\hat{b}^\dagger_{\bf q\nu}\hat{b}_{\bf q\nu}+\frac{1}{2}\Big)}\\
        &&+\frac{1}{\sqrt{N}}\sum_{\bf q\nu}\sum_{m,n\bf k\sigma}
        \hat{c}^\dagger_{m\bf k+q\sigma}\hat{c}_{n\bf k\sigma}
        \Big(\hat{b}^\dagger_{\bf -q\nu}+\hat{b}_{\bf q\nu}\Big)
        g^{\nu \bf q}_{m{\bf k+q},n{\bf k}},\nonumber
\en
the parameters of which are set to the 
\textit{ab initio} computed values. In the above $m,n$ are band
labels, $\sigma$ is a spin label, $\bf k,q$ are wave vectors 
in the 1BZ, $\{\hat{c},\hat{c}^\dagger\}$
and $\{\hat{b},\hat{b}^\dagger\}$ are electron and phonon ladder operators, 
$\{\epsilon\}$ are electronic eigenvalues
and $\{g\}$ describe the strength of the scattering between electrons and phonons.
It is standard to set the electronic energies to the Kohn-Sham eigenvalues,
the phonon frequencies to the Born-Oppenheimer frequencies and to 
extract the values of the $g$ parameters from the deformation potential.
Standard field theory methods\cite{Mahan} are then employed to derive 
the phonon linewidths, $\gamma_\nu$, from this Hamiltonian.

\section{Phonon anharmonicity}
\label{section phonon}
The position operator for the ions in a crystal can be
represented as
\be
	\hat{\bf r}_{\kappa}({\bf R}) &=& {\bf R}+{\bf b}_\kappa
				+\hat{\bf u}_{\kappa}({\bf R}),
\en
where ${\bf R}$ is a lattice vector, $\bf b_\kappa$ is the basis
vector for ion $\kappa$ and  $\hat{\bf u}_{\kappa}({\bf R})$ is the operator
representing the displacement of the ion from its equilibrium
position. Within the adiabatic approximation, which assumes the electronic
system instantaneously adapts to the ionic positions, the total energy
as a function of the ionic positions can be taken as an effective
potential for the ions which thus dictates their dynamics. This potential
is expressed as
\be
\label{total potential}
	\hat{U}[\{\hat{\bf u}\}] &=& U_0+\sum_{n=2}^{\infty}\hat{U}_n[\{\hat{\bf u}\}],
\en
with
\be
	\hat{U}_n[\{\hat{\bf u}\}] &=&
		 \frac{1}{n!}
		\sum_{\{\alpha\kappa\bf R\}}
		\hat{u}_{\kappa_1}^{\alpha_1}({\bf R}_1)...
		\hat{u}_{\kappa_n}^{\alpha_n}({\bf R}_n)
		\nonumber\\
		&&\hspace{2cm}\times\Phi_{\kappa_1...\kappa_n}^{\alpha_1...\alpha_n}
			({\bf R}_{1},...,{\bf R}_{n}),
\en
where the Greek symbols $\alpha_1,...,\alpha_n$ represent cartesian coordinates.
It is assumed that the crystal is stable and that, consequently,
the linear term in the displacements vanishes identically. The dynamics
of the ionic degrees of freedom are then described by the effective Hamiltonian
\be
	\hat{H} &=& \hat{T}+\hat{U},
\en
where $\hat T$ is the kinetic energy operator of the ions.
It is convenient to consider a canonical change of variable to 
reciprocal space of the form
\be
	\hat{\bf u}_{\kappa}({\bf R})	&=& \frac{1}{\sqrt{N}}
			\sum_{\bf q} e^{i{\bf q\cdot R}}
				\hat{\bf u}_{\kappa}({\bf q})
\en
In terms of these new position-like variables, the potential terms can be
expressed as
\be
\label{anharmonic potential}
	\hat{U}_n[\{\hat{\bf u}\}] &=&
		 \frac{1}{n!}\frac{1}{N^{\frac{n}{2}-1}}
		\sum_{\{\alpha\kappa \bf q\}}
		\hat{u}_{\kappa_1}^{\alpha_1}({\bf q}_1)...
		\hat{u}_{\kappa_n}^{\alpha_n}({\bf q}_n)\nonumber\\
		&&\hspace{1cm}\times \Phi_{\kappa_1...\kappa_n}^{\alpha_1...\alpha_n}
			({\bf -q}_1,...,{\bf -q}_n);
\en
some useful symmetry relations pertaining to these anharmonic coefficients
are reminded in appendix \ref{anharmonic appendix}.

\subsection{Harmonic phonons}
If the potential energy expansion
is truncated after the second order, the resulting approximate Hamiltonian
is harmonic, and leads to the standard small oscillations problem. In this 
case:
\be
	\hat{U}_2 &=&
		 \frac{1}{2}
		\sum_{\{\alpha\kappa\}}
		\sum_{{\bf q}}
		\Big[\hat{u}_{\kappa_1}^{\alpha_1}({\bf q})\Big]^\dagger
		D_{\kappa_1\kappa_2}^{\alpha_1\alpha_2}({\bf q})
		\hat{u}_{\kappa_2}^{\alpha_2}({\bf q}),
\en
where the usual dynamical matrix has been defined as
\be
	D_{\kappa_1\kappa_2}^{\alpha_1\alpha_2}({\bf q}) &\equiv&
		\Phi_{\kappa_1\kappa_2}^{\alpha_1\alpha_2}
			({\bf q},{\bf -q}).
\en
It is standard to consider a canonical transformation to ladder operators
of the form
\be
	\hat{u}_{\kappa}^{\alpha}({\bf q}) =
		\sum_{\nu} 
		x_{\kappa\nu}^{\alpha}({\bf q})
		\hat{A}_{\bf q\nu};\hspace{5mm}
		\hat{A}_{\bf q\nu}= \hat{b}_{\bf q\nu}+
		\hat{b}^\dagger_{\bf -q\nu}.
\en
The displacement vectors are defined as
\be
		{\bf x}_{\kappa\nu}({\bf q})
		&=&
		\sqrt{\frac{\hbar}{2M_\kappa\omega_\nu({\bf q})}}
		{\bf E}_{\kappa\nu}({\bf q}),
\en
where $M_\kappa$ is the mass of ion $\kappa$ and the polarization vectors 
$\bf E$, which are chosen to be orthonormal, are solutions of the hermitian 
eigenvalue problem
\be
	\sum_{\kappa_2\alpha_2}
		\frac{D_{\kappa_1\kappa_2}^{\alpha_1\alpha_2}({\bf q})}
		{\sqrt{M_{\kappa_1}M_{\kappa_2}}}
		E_{\kappa_2\nu}^{\alpha_2}({\bf q})
		&=& \omega_\nu({\bf q})^2
		E_{\kappa_1\nu}^{\alpha_1}({\bf q}),
\en
which also yields the harmonic phonon frequencies. It is useful to define
a "mode mass" $M_\nu$ as
\be
	M_\nu({\bf q})
		&=&
	\frac{ \sum_\kappa M_\kappa | {\bf x}_{\kappa\nu}({\bf q})|^2}
		{\sum_\kappa | {\bf x}_{\kappa\nu}({\bf q})|^2};
\en
this quantity is then a gauge of what type of ions are involved in a given mode.
The harmonic part of the ionic Hamiltonian can finally be expressed as 
\be
	\hat{H}_h &=& \hat{T}+\hat{U}_2,\\
		  &=& \sum_{\bf q,\nu} \hbar \omega_\nu({\bf q})
			\Big(\hat{b}_{\bf q\nu}^\dagger\hat{b}_{\bf q\nu}
			+\frac{1}{2}\Big).
\en

\subsection{Phonon-phonon interaction}
\begin{figure}
\subfigure[Loop]   {\includegraphics{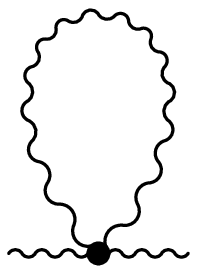}}
\subfigure[Bubble] {\includegraphics{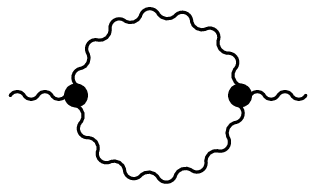}}
\subfigure[Tadpole]{\includegraphics{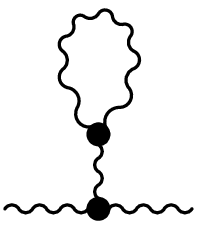}}
\caption{\label{diagrams} Lowest order anharmonic self-energy diagrams. 
The lines represent phonon propagators and the vertices third order ("bubble"
and "tadpole") and fourth order ("loop") anharmonic coupling.} 
\end{figure}

When the anharmonic contributions to the potential cannot be neglected,
 the ionic Hamiltonian can be expressed in terms
of the harmonic Hamiltonian plus phonon-phonon interaction terms.
The anharmonic coefficients can be expressed in terms of the harmonic basis,
and lowest order contributions to the self-energy anharmonic correction 
for the mode $(\bf q\nu)$ are given by\cite{PhysRev.128.2589} 
\begin{widetext}
\be
\label{pi loop}
\Pi^{(L)}_{\nu}({\bf q},\omega)
		&=&
		\frac{1}{2N}\sum_{\nu_1,{\bf q}_1}
		\Phi_{\nu,\nu,\nu_1,\nu_1}({\bf q},{\bf -q},{\bf q}_1,{\bf -q}_1)
		\Bigg(2n_B\Big(\hbar\omega_{\nu_1}({\bf q}_1)\Big)+1\Bigg),\\
\Pi^{(T)}_{\nu}({\bf q},\omega)
		&=&
	-\frac{1}{N}\sum_{{\bf q}_1}\sum_{\nu_1 \nu_2}
		\Phi_{\nu_1,\nu_1,\nu_2}({\bf -q}_1,{\bf q}_1,{\bf 0})
		\Phi_{\nu_2,\nu,\nu}({\bf 0},{\bf q},{\bf -q})
		\frac{2n_B\Big(\hbar\omega_{\nu_1}({\bf q}_1)\Big)+1}
		{\hbar\omega_{\nu_2}({\bf 0})},\\
	\Pi^{(B)}_{\nu}({\bf q},\omega)
		&=&
		-\frac{1}{2N}
	\sum_{{\bf q}_1,{\bf q}_2}\sum_{\nu_1 \nu_2}
	\sum_{\bf G}\delta_{{\bf q}_1+{\bf q}_2+{\bf q},{\bf G}}
		|\Phi_{\nu,\nu_1,\nu_2}({\bf q},{\bf q}_1,{\bf q}_2)|^2
	F\Big(\omega,\omega_{\nu_1}({\bf q}_1),\omega_{\nu_2}({\bf q}_2)\Big),
\en
where
\be
	F(\omega,\omega_1,\omega_2)	
	&=&
	\frac{1}{\hbar}\Bigg[\frac{2 \Big(\omega_1+\omega_2\Big)
	\Big(1+n_B(\omega_1)+n_B(\omega_2)\Big)}
	{\Big(\omega_1+\omega_2\Big)^2-\Big(\omega+i\delta\Big)^2}
	+\frac{2 \Big(\omega_1-\omega_2\Big)\Big(n_B(\omega_2)-n_B(\omega_1)\Big)}
	{\Big(\omega_2-\omega_1\Big)^2-\Big(\omega+i\delta\Big)^2}\Bigg]
\en
and
\be
	\Phi_{\nu_1...\nu_n}({\bf q}_{1},...,{\bf q}_{n}) &=&
		\sum_{\{\alpha\kappa\}}
		\Phi_{\kappa_1...\kappa_n}^{\alpha_1...\alpha_n}
		({\bf q}_{1},...,{\bf q}_{n})
		x_{\kappa_1\nu_1}^{\alpha_1}(-{\bf q}_1)...
		x_{\kappa_n\nu_n}^{\alpha_n}(-{\bf q}_n).
\en
\end{widetext}
In the above, the quantity $n_B$ refers to the usual bosonic occupation factor.
The labels $(T),(L)$ and $(B)$ refer to the "tadpole", "loop" and "bubble" diagrams,
as schematically represented in Figure \ref{diagrams}. It is noteworthy that
only the "bubble" contribution actually depends on the frequency $\omega$, 
and that only this term will have an imaginary contribution.
Furthermore, the "tadpole" diagram vanishes by symmetry in this system. 
The results above will be specialized to the point 
${\bf q}_X = \pi/a(0,0,1)$ on the side of the zone, at zero temperature.

\section{Computational details}
\label{section computations}
Electronic properties were computed using density functional theory
(DFT) 
as implemented in the {\sc Quantum-ESPRESSO} package\cite{QE-2009}. 
The exchange-correlation was treated using the  
generalized gradient approximation (GGA) of Perdew, Burke and Ernzerhof (PBE)
\cite{PhysRevLett.78.1396,PhysRevLett.77.3865}. 
Ultrasoft pseudopotentials\cite{PhysRevB.41.7892} were used, where 
3$s$ and 3$p$ states of aluminium were treated as valence. The plane-wave
basis cutoff was set to 80 Ry.  First Brillouin Zone (1BZ) integrations were
performed as sums on a $24\times24\times24$ 
Monkhorst-Pack 
$\bf k$-mesh, using a smearing parameter of 20\thinspace mRy.
Phonon properties were computed using density-functional perturbation theory (DFPT)
\cite{dfpt,QE-2009}. 
Interatomic force constants (IFC) were obtained 
from dynamical matrices computed on a $12\times12\times12$ 
$\bf q$-mesh. 
The electron-phonon coupling computations were performed using electronic
and phonon quantities interpolated on a fine $72\times72\times72$ mesh.

\section{Results}
\label{section results}
\begin{figure*}
\includegraphics[width=1.0\textwidth,angle=0]{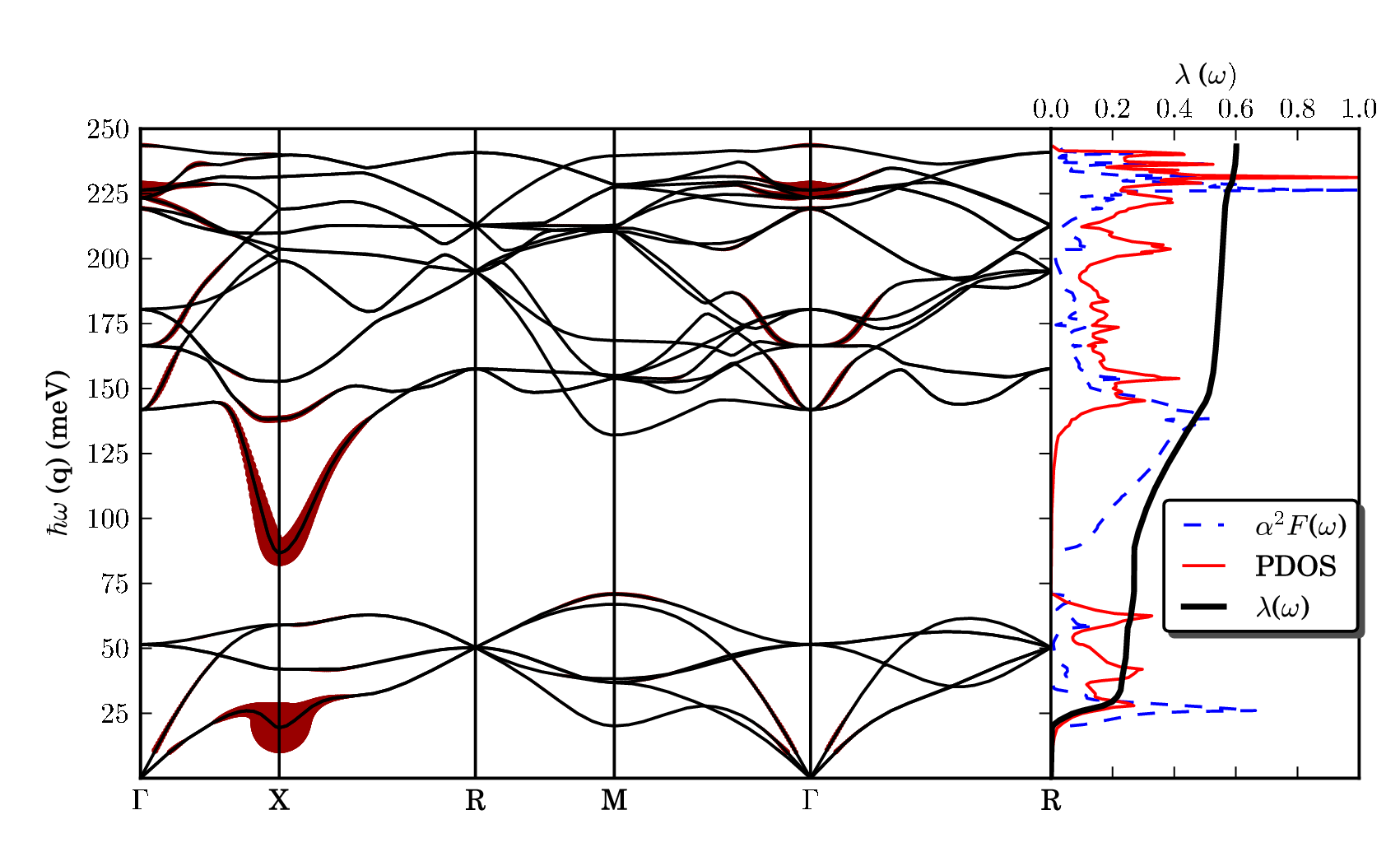}
\caption{\label{ph}(Color online) (Left panel) Phonon dispersion relation computed 
using DFPT.
The areas of the red circles overlapping the phonon dispersion are
proportional to $\lambda_{\nu{\bf q}}$.  It is clear that the bulk of the 
electron phonon coupling can be attributed to modes in a region near X with frequencies
$\sim$\thinspace 20\thinspace meV(X$_1$) and $\sim$\thinspace 85\thinspace meV(X$_2$).
(Right panel)
Eliashberg spectral function and phonon density of states PDOS (both in arbitrary
units), as well as partially integrated value of $\lambda$.
}
\end{figure*}

The phonon dispersion was calculated for the $Pm\bar 3n$ structure with
a lattice constant $a = 5.82$\thinspace$a_0$, which yielded a computed pressure
of 109\thinspace GPa. The phonon spectral function $\alpha^2F$, as well as
the electron-phonon coupling parameter $\lambda$ were also computed; results
can be seen in Fig. \ref{ph}. 
The values we have obtained are $\lambda\simeq 0.61$ and 
$\hbar \omega_{log} \simeq 68$\thinspace meV, leading
to 12\thinspace K $\lesssim$ T$_c$ $\lesssim$ 19\thinspace K
($0.14\geq\mu^*\geq0.1$).
  As is clear from Fig. \ref{ph}, the bulk of the
contribution to $\lambda$ comes from narrow regions in the 1BZ centered at
X, for modes at $\sim 20$\thinspace meV and $\sim85$\thinspace meV;
 attention has been focused on the relevant modes at X, 
assuming that they are representative of modes in that region of the 1BZ.
These doubly degenerate modes will henceforth be referred to as X$_1$ (20 meV) 
and X$_2$ (85 meV) and will be labelled as $\nu_X$.
Interestingly, for these modes the motion of hydrogen ions is
perpendicular to their chains; the displacements for X$_2$ are mostly
that of hydrogen ions (with a mode mass of 1.1 $M_H$), 
whereas the displacements of X$_1$ involve both types of ions (mode mass of 5.7 $M_H$). 

\begin{figure}
\centering
        \includegraphics{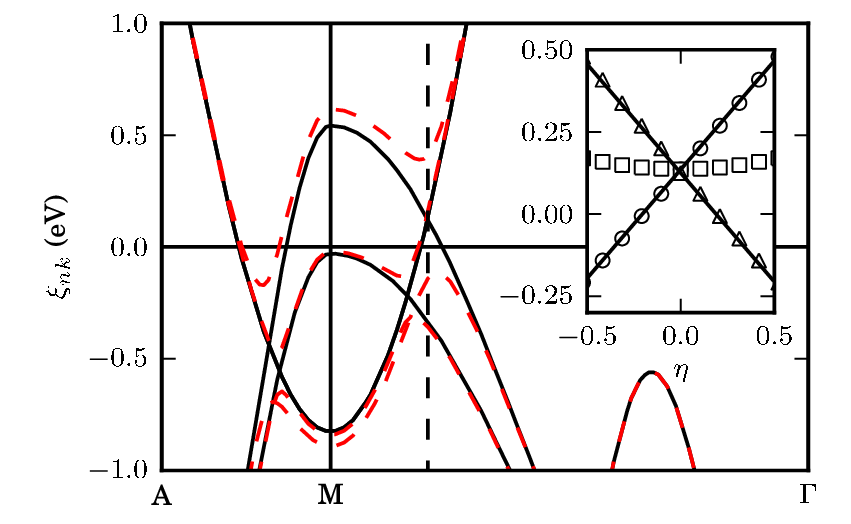}
  \caption{\label{fig:split_bands}
        (Color online)
        Electronic bands near the Fermi energy in the tetragonal supercell.
        The bands for the unperturbed structure are shown (full black line), 
        as well as for a X$_2$ frozen phonon displacement corresponding to 
        $\eta\simeq 0.4$ (dashed red line). The zero of energy is set
	at the Fermi energy of the unperturbed system. The X$_1$ data is not
        significantly different. The dashed vertical
        line indicates the point ${\bf k}_0$ where the unperturbed bands cross.
        (Inset) Perturbed eigenvalues as a function of $\eta$ corresponding
        to the degenerate energy indicated by the dashed line. 
        The splitting is linear with the distortion for two bands (circles and triangles)
        and remains quadratic for the third (squares); 
        the value of $|g|:$ is directly related to the slope of the change in band
        energy with $\eta$, and is equal to $\sim$\thinspace 470\thinspace meV
	in this case.  }
\end{figure}

\begin{figure*}
\subfigure[\; X$_1$]{
	\includegraphics[width=0.45\textwidth,angle=0]{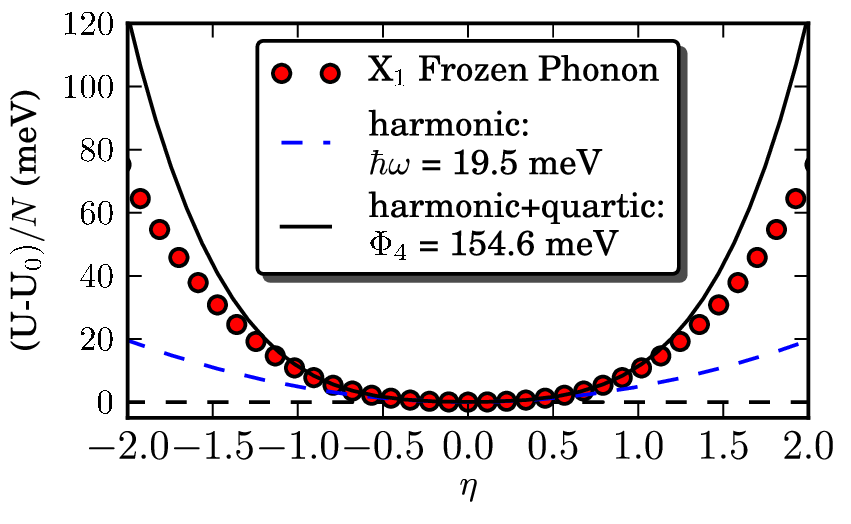}}
\subfigure[\; X$_2$]{
	\includegraphics[width=0.45\textwidth,angle=0]{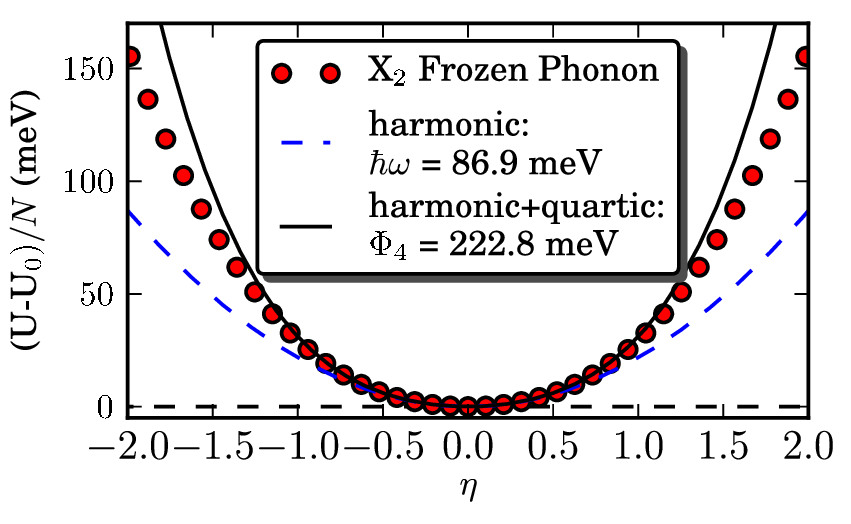}}
\caption{\label{FP} (Color online) Frozen phonon total energy calculations for the
modes at X with frequency $\sim$\thinspace 20\thinspace meV (X$_1$) and 
$\sim$\thinspace 85\thinspace meV (X$_2$), as a function of a displacement
parameter $\eta$. The values of the parameters $\hbar\omega$ and
$\Phi_4$ are obtained from finite difference schemes. 
}
\end{figure*}
To better understand the origin of the large contribution to $\lambda$ 
of modes around X, frozen phonon-perturbed bands were computed for X$_1$ and X$_2$. 
A displacement of the form
\be
\label{eq:displacement}
	{\bf u}_\kappa({\bf R};\eta) &=& \eta  e^{i{\bf q}_X\cdot {\bf R}}
					{\bf x}_{\kappa\nu_X}({\bf q}_X) 
\en 
was imposed on the ions in a supercell geometry for various
values of the unitless parameter 
\footnote{ To get a sense of scale, note that, for the ion most
displaced, 
$|{\bf x}_{\kappa\nu_x}|\simeq0.15\thinspace  a_0 \simeq 0.03\thinspace a$ for X$_1$
and 
$|{\bf x}_{\kappa\nu_x}|\simeq0.14\thinspace  a_0 \simeq 0.02\thinspace a$ for X$_2$.}
$\eta$
 (the displacement can be chosen
to be real because of the symmetry of the $\bf q_X$ point). 
The appropriate  supercell corresponds to a doubling of the original cell 
along the $c$ axis, yielding a tetragonal cell.
In the presence of this frozen-in perturbation,
the Kohn Sham Hamiltonian can be expressed to first order in $\eta$ as
\be
\hat{h}_{KS} &\simeq&
        \sum_{n\bf k\sigma}\Big[ \epsilon_{n\bf k}
        \hat{c}^\dagger_{n\bf k\sigma}\hat{c}_{n\bf k\sigma}\\
        &&\hspace{1cm}+\eta \sum_{m}
        \hat{c}^\dagger_{m{\bf k+q}_X\sigma}\hat{c}_{n\bf k\sigma}
        g^{\nu_X {\bf q}_X}_{m{\bf k+q}_X,n{\bf k}}\Big].\nonumber
\en
The correction to the eigenenergies $\epsilon_{n\bf k}$ will be of order $\eta^2$
at a generic $\bf k$ point.  However, in the case of band degeneracy,
the usual response formalism breaks down and 
the change in the band energy can be linear in $\eta$.
The unperturbed cubic system has Fermi sheets
centered at the R (doubly degenerate) and M (non degenerate)  points.
As can be seen in Fig. \ref{fig:split_bands},
these two Fermi sheets are centered about the M point in the supercell geometry.
The two sets of bands intersect close to the Fermi energy along the $\Gamma$-M
direction at a point ${\bf k}_0$; at this point, the Kohn Sham Hamiltonian
can be modeled, to linear order in $\eta$, as
\be
	{\bf h}_{KS} &=& \left(
			\begin{array}{ccc}
			\epsilon_0 & 0 & \eta g\\
			0 & \epsilon_0 & \eta g\\
			\eta g^* & \eta g^* & \epsilon_0
			\end{array} \right),
\en
where
\be
	\epsilon_0 \equiv \epsilon_{n {\bf k}_0};\hspace{5mm}
	g \equiv g_{m {{\bf k}_0+\bf q_X},n{\bf k}_0}^{\nu_X \bf q_X}.
\en
The band label indicates one of the three $\eta=0$ degenerate states,
and it is assumed that the only relevant coupling to linear order 
is between the non-degenerate M band and the doubly degenerate R bands.
The eigenvalues of this Hamiltonian matrix are given by
\be
	\epsilon &=& 0, \pm \sqrt{2}\eta |g|
\en
and it is straightforward to extract a value for $|g|$.
We find that $|g| \simeq 440$\thinspace meV for X$_1$ and
 $|g| \simeq 470$\thinspace meV for X$_2$.
These couplings
are very large on the phonon energy scale, and suggest that the 
origin of the large linewidths lies in strong scattering between the Fermi sheets
mentioned above.
The large coupling associated to X$_1$ and X$_2$
prompted further investigation of these modes.

In order to gauge the anharmonicity of these modes, total energy frozen phonon 
calculations were also performed. A displacement of the form
given by Eq. (\ref{eq:displacement}) corresponds to 
\be
	u_\kappa^\alpha({\bf q};\eta) &=& \eta \sqrt{N}
					x_{\kappa\nu_X}^{\alpha}({\bf q}_X) 
					\delta_{\bf q,q_X}.
\en 
 It is important to note that, for a finite value of $\eta$,
this does not correspond to a realistic configuration of the ions. Indeed, 
the zero-point energy of a mode is in the order of meV, an energy that must
be shared by all ions (a number of order $N$). Thus, for a single mode,
only an infinitesimal amount of energy can be assigned to any ion, leading
to infinitesimal average displacement. It is then the sum on all modes that yield
a finite average displacement for any ion. This discussion does not invalidate
the frozen phonon calculations, as they are only performed to extract 
anharmonic coefficients.

According to the expressions for the anharmonic energy as a function of
displacement, Eq. (\ref{total potential}) and (\ref{anharmonic potential}),
the total energy per unit cell for a given displacement should then be
\be
	\frac{U[\eta]}{N} = 
	\frac{U_0}{N} 
	+\frac{\eta^2}{4}\hbar\omega_{\nu_X}({\bf q}_X)
	+\frac{\eta^4}{24} \Phi_{4,\nu_X} +O(\eta^6)
\en
where
\be
	\Phi_{4,\nu_X} &=&\Phi_{\nu_X\nu_X\nu_X\nu_X}({\bf q}_X,{\bf q}_X,{\bf q}_X,{\bf q}_X)
\en
has been defined for convenience (note that ${\bf q}_X$ and ${\bf -q}_X$ are 
equivalent points by reciprocal lattice periodicity).
 It is straightforward to extract the values
of $\hbar\omega_{\nu_X}({\bf q}_X)$ and $\Phi_{4,\nu_X}$ 
from the energy as a function of $\eta$
using a finite difference scheme; results can be seen in figure \ref{FP}.
As can be seen on the figure, the quartic contribution to the potential is
very large, with $\Phi_4\simeq 155$\thinspace meV for X$_1$ and
$\Phi_4\simeq 223$\thinspace meV for X$_2$. In the \textit{naive} approximation
that the quartic anharmonic coupling is constant throughout the zone and
that coupling to other modes can be neglected,
namely
\be
\lefteqn{
   \Phi_{\nu_X,\nu_X,\nu,\nu}({\bf q}_X,{\bf q}_X,{\bf q},{\bf -q})
	\simeq}\nonumber\\
	&&\hspace{1cm}\delta_{\nu,\nu_X}
   \Phi_{\nu_X,\nu_X,\nu_X,\nu_X}({\bf q}_X,{\bf q}_X,{\bf q}_X,{\bf q}_X),
\en
 and at zero temperature,
this leads to a frequency renormalization through the "loop" diagram
of 77\thinspace meV for X$_1$ and 111\thinspace meV for X$_2$
(see equation \ref{pi loop}).

Drawing conclusions from results at a single $\bf q$ point can be 
premature, as is examplified by the case of MgB$_2$:
frozen phonon calculations similar to those presented above
suggested that the $E_{2g}$ modes at $\Gamma$ should be 
highly anharmonic\cite{PhysRevLett.87.037001}; refined calculations
of the "loop" and "bubble" diagrams for this mode revealed that in fact
this is not the case\cite{PhysRevB.68.220509,d'astuto:174508}.
The point is that a frozen phonon calculation provides no information on
the cubic coupling, which can largely cancel the quartic contribution,
nor does it account for the fact that the anharmonic coefficients have
dispersions (\textit{ie} are functions of $\bf q$).
Nevertheless, the unusually large value of $\Phi_4$ prompted 
a more thorough investigation of anharmonicity for these modes.

\begin{table}
\begin{center}
\begin{tabular}{ c d d c d d }
\hline
\hline
 & \multicolumn{2}{c}{$\hbar\omega_{\nu}({\bf q}_X)$}
 & \hspace{5mm}
 & \multicolumn{2}{c}{ $\Phi_{4,\nu_X}$}
\\
 & \multicolumn{1}{c}{\mbox{FP}}
 & \multicolumn{1}{c}{\mbox{DFPT}}
 &
 & \multicolumn{1}{c}{\mbox{FP}}
 & \multicolumn{1}{c}{\mbox{SFD}}
 \\
		\hline
 $X_1$ & 19.5 & 19.4 && 154.6 & 154.9\\
 $X_2$ & 86.9 & 86.8 && 222.8 & 224.2\\
		\hline
		\hline
\end{tabular}
\end{center}
\caption{\label{table_phi4} Comparing harmonic frequencies and $\Phi_{4}$ parameters 
(in meV) obtained by different methods. DFPT means ``density functional perturbation theory", FP 
means ``frozen phonon" and SFD means ``supercell finite difference".
Results are seen to agree very well between different calculations, giving confidence
on the convergence of the SFD method.  }
\end{table}

\begin{table}
\begin{center}
\begin{tabular}{ddddddd}
\hline \hline
      & \multicolumn{1}{c}{$\Pi^{(L)}_{\nu}$}
      & \multicolumn{1}{c}{$\Pi^{(B)}_{\nu}$} 
      & \multicolumn{1}{c}{$\Pi_{\nu}$}
      & \multicolumn{1}{c}{$\hbar\omega_{\nu}$}
      & \multicolumn{1}{c}{$\hbar\Omega_{\nu}$} 
      & \multicolumn{1}{c}{shift (\%)}\\
\hline
   X_1 &              22.9             
       &              -6.3             
       &              16.6             
       &              19.4             
       &              31.9             
       &              65               \\
   X_2 &              37.3             
       &              -12.0             
       &              24.9             
       &              86.8             
       &              108.9
       &              25               \\
\hline \hline
\end{tabular}
\end{center}
\caption{\label{table_Omega} 
Computed values of the self-energy corrections for
modes X$_1$ and X$_2$, at zero temperature and zero frequency, 
and comparison of the renormalized 
frequency, $\Omega_{\nu}$, with the harmonic frequency $\omega_{\nu}$.
 All energies are
in meV.  The relative shift is seen to be quite large.}
\end{table}

The necessary coefficients were obtained by
finite differencing of dynamical matrices computed for appropriate supercells
(henceforth the  ``supercell finite difference", of SFD,  method);
a complete discussion of the formalism can be found in appendix
\ref{finite difference appendix}.
Briefly, anharmonic parameters were obtained by computing dynamical matrices with
ions slightly displaced according to the polarizations of modes X$_1$ and X$_2$.
A centered, five points finite
difference scheme was applied to these dynamical matrices, 
yielding partially mode-projected anharmonic coefficients of the form 
$\Phi^{\alpha_1\alpha_2}_{\kappa_1\kappa_2;\nu_X}({\bf q})$
and
$\Phi^{\alpha_1\alpha_2}_{\kappa_1\kappa_2;\nu_X\nu_X}({\bf q})$.
These coefficients were obtained on a $4\times4\times4$ $\bf q$-mesh; 
Fourier interpolation was used to approximate them throughout the 1BZ.

The frequency dependent self-energy was computed in a range of interest, 
and the phonon spectral function was obtained\cite{Mahan}.
In particular, by using the Lehmann representation\cite{Mahan}, it can be shown
that
\be
        \int_0^\infty d\omega \frac{ B_{\nu}({\bf q},\omega)}{\omega}
        &=& \frac{ \omega_{\nu}({\bf q})} {\Omega_{\nu}({\bf q},0)^2},
\en
where 
\be
        \Big(\hbar\Omega_{\nu}({\bf q},\omega)\Big)^2 &=&
        \Big(\hbar\omega_{\nu}({\bf q})\Big)^2
        +2\hbar\omega_{\nu} \Pi_{\nu}({\bf q},\omega).
\en
The computed values of the ``loop" and ``bubble" self energy diagrams,
at zero temperature and zero frequency, can be seen in 
Table \ref{table_Omega}, along with the values of $\Omega_{\nu}({\bf q}_X,0)$.
The ``loop" contributions are quite large, but 3 to 4 times 
smaller than what the \textit{naive} estimate based on dipsersionless parameters
suggested. The cubic terms, which are real and negative at zero frequency,
further reduce the estimate of the total self energy. The latter
remains large
enough however to strongly renormalize $\Omega_\nu$ with respect to $\omega_\nu$,
yielding
a sizeable relative shift of 65 \% ( 25 \%) of the frequency of mode X$_1$ (X$_2$).

The contributions to the linewidths coming from anharmonic effects were also computed
at 0 and 300\thinspace K, for a \textit{fixed} lattice geometry 
(a fixed value of $a$).
Although a proper analysis of  the temperature dependence should include 
the lattice expansion (which is beyond the scope of this work), 
we expect that the fixed cell calculations should still be indicative
of the behavior of the linewidths at fixed pressure.
The widths were found to be vanishingly small (0.3 meV) for mode X$_1$
and 0.6 meV (4.6\thinspace meV) for mode X$_2$ at 0\thinspace K (300\thinspace K). 
Given that the electron-phonon contribution to the linewidth at
X$_2$ is about 10\thinspace meV, it should be possible to observe the
temperature dependent contribution to this mode's linewidth, providing
a possible experimental signature of the effects described here.

In the approximation that anharmonicity does not affect the electron-phonon coupling
(a reasonable assumption given that the $g$ parameters are obtained from the
deformation potential method, which is not affected by anharmonicity), 
the mode coupling is renormalized by anharmonicity and becomes
\be
        \lambda^{(anh)}_{\bf q\nu} &=&\frac{1}{\pi\hbar N(0)}
                \frac{\gamma_\nu({\bf q})}{\Omega_\nu({\bf q},0)^2}.
\en
A complete calculation of the renormalized value $\lambda^{(anh)}$
from equation (\ref{lambda sum})
would require knowledge of the anharmonic coefficients at points other 
than $\bf q_X$ and is beyond the scope of this work.
However we can estimate an upper bound for the effect of the anharmonicity on the 
electron-phonon coupling. 
From Fig. \ref{ph} it is reasonable to assume that the electron-phonon coupling
from 0 to $\sim$\thinspace 35\thinspace meV can be attributed to the region near X$_1$ 
(partial value $\lambda_1 \simeq 0.23$) and 
that the electron-phonon coupling from $\sim$\thinspace 85\thinspace meV 
to $\sim$\thinspace 135\thinspace meV 
can be attributed to the region near X$_2$ (partial value $\lambda_2 \simeq 0.2$); 
the rest of the coupling is lumped together and assumed unaffected by anharmonicity 
(partial value $\lambda_3 = \lambda-\lambda_1-\lambda_2\simeq 0.18$).
By estimating that the harmonic Eliashberg function is composed of two properly
normalized $\delta$ peaks at $\omega_1$ and $\omega_2$ (the frequencies corresponding to X$_1$ 
and X$_2$) plus features
unaffected by anharmonicity away from these frequencies, 
the renormalized coupling is given by
\be
        \lambda^{(anh)} &\simeq& \Big(\frac{\omega_{1}}{\Omega_{1}}\Big)^2 \lambda_1
                               +\Big(\frac{\omega_{2}}{\Omega_{2}}\Big)^2 \lambda_2
                               +\lambda_3
                        = \;0.39,
\en
where $\Omega_1$ and $\Omega_2$ are the renormalized frequencies of X$_1$ and X$_2$.
A similar treatment yield $\hbar \omega^{(anh)}_{log}\simeq 125$\thinspace meV;
using these renormalized parameters, the Allen-Dynes modification to the
McMillan formula yields 1.5\thinspace K $\lesssim$ T$_c$ $\lesssim$ 5\thinspace K
for $0.14\geq\mu^*\geq0.1$, suggesting that the anharmonic renormalization
of the phonon spectrum, which acts to stiffen the modes which provides
most of the contribution to $\lambda$, leads to a strong reduction 
on T$_c$ compared to the results obtained from the harmonic phonon spectrum.

\section{Conclusion}

In this work, a full analysis of anharmonic effects to lowest order 
has been presented for modes in a region of the 1BZ which provides most
of the contribution to $\lambda$, and this analysis
has been compared to \textit{naive} frozen phonon calculations.
It has been shown that the modes X$_1$ and X$_2$ of AlH$_3$ are strongly 
renormalized by anharmonicity,
although less so than a frozen phonon calculation might have suggested.
Indeed, it is found that the frequency of X$_1$ is renormalized to 31.9\thinspace meV
from 19.4\thinspace meV (a 65\% shift)
and that the frequency of X$_2$ is renormalized to 108.9\thinspace meV
from 86.8\thinspace meV (a 25\% shift).
Furthermore, it is expected that anharmonicity induces a large, temperature 
dependent contribution to the X$_2$ mode linewidth, which could provide 
an experimental signature of the effect. 
A rough estimate suggests that renormalization could lead to a great reduction of 
the computed value of T$_c$, which could be as low as 2\thinspace K, compared
to the harmonic prediction of T$_c$$\simeq$20\thinspace K.

Thus, anharmonicity in AlH$_3$ might play a role in explaining why the measured 
and computed superconducting transition temperatures are in qualitative disagreement.
There could be further phonon frequency renormalization due to 
the large electron-phonon coupling and, 
given that the electronic bandwidth near the Fermi energy is quite small, 
non-adiabatic effects in this system could also be substantial.

\acknowledgments{
We are grateful to I. Errea, J.M. Perez-Mato, 
and N.W. Ashcroft for fruitful discussions. 
We acknowledge financial support from UPV/EHU (Grant No. IT-366-07).
}

\appendix
\section{Symmetries of anharmonic coefficients}
\label{anharmonic appendix}
The anharmonic coefficients can be defined as
\be
	\Phi_{\kappa_1...\kappa_n}^{\alpha_1...\alpha_n}
			({\bf R}_{1},...,{\bf R}_{n}) &=&
			\frac{\partial^n U[\{\bf u\}]}
	{\partial u^{\alpha_1}_{\kappa_1}({\bf R}_1)...
	\partial u^{\alpha_n}_{\kappa_n}({\bf R}_n)}\Bigg|_{\bf u = 0}
\en
which immediately implies that they are real and symmetric under permutations of
indices. Define the Fourier transformed coefficients as
\be
	\Phi_{\kappa_1...\kappa_n}^{\alpha_1...\alpha_n}
			({\bf q}_{1},...,{\bf q}_{n}) &=&
		 \frac{1}{N}
		\sum_{\{\bf R\}}
		e^{-i({\bf q}_1\cdot{\bf R}_1+...+{\bf q}_n\cdot{\bf R}_n)}\nonumber\\
		&&\hspace{0.5cm}\times
		\Phi_{\kappa_1...\kappa_n}^{\alpha_1...\alpha_n}
			({\bf R}_{1},...,{\bf R}_{n}).
\en
Translational symmetry, which can be expressed as 
\be
	\lefteqn{\Phi_{\kappa_1...\kappa_n}^{\alpha_1...\alpha_n}
			({\bf R}_{1},...,{\bf R}_{n}) =}\nonumber\\
	&&\hspace{1cm}\Phi_{\kappa_1...\kappa_n}^{\alpha_1...\alpha_n}
			({\bf R}_{1}+{\bf R},...,{\bf R}_{n}+{\bf R})
\en
for all lattice vectors $\bf R$, implies that 
$\Phi_{\kappa_1...\kappa_n}^{\alpha_1...\alpha_n}
({\bf q}_{1},...,{\bf q}_{n})$ vanishes unless ${\bf q}_1+...+{\bf q}_n$ is 
a reciprocal lattice vector. Furthermore,
\be
	\Phi_{\kappa_1...\kappa_n}^{\alpha_1...\alpha_n}
			({\bf q}_{1},...,{\bf q}_{n}) &=&
	\Phi_{\kappa_1...\kappa_n}^{\alpha_1...\alpha_n}
			({\bf q}_{1}+{\bf G}_1,...,{\bf q}_{n}+{\bf G}_n)\nonumber\\
\en
for any set of reciprocal lattice vectors $\{\bf G\}$.

\section{Extracting anharmonic parameters from dynamical matrices}
\label{finite difference appendix}
The lowest order anharmonic correction to the self energy,
\be
\Pi &=&
		\Pi^{(L)}+
		\Pi^{(B)}+
		\Pi^{(T)},
\en
involves anharmonic coefficients of third ("bubble" and "tadpole") and
fourth ("loop") order. The most efficient and elegant way of obtaining
these parameters is the use of the $2n+1$ theorem within the context
of density-functional perturbation theory\cite{dfpt}. Briefly, this
theorem guarantees that derivatives of the total energy up to order
$2n+1$ can be obtained from knowledge of the derivatives of the 
wave-functions to order $n$. In practice, however, only the 
first order derivatives of the wavefunctions (\textit{ie} $n=1$) are
readily available, and finite-difference schemes are employed to
obtain fourth order coefficients.

An alternative way of obtaining the necessary coefficients is through 
the frozen phonon method and finite differencing. What this method
lacks in elegance, it makes up for by its simplicity and straightforward
use, without the need for specialized software.  The frozen phonon approach 
for this purpose
is impractical for arbitrary $\bf q$ in the 1BZ, as it implies computations
with potentially very large supercells. However, the interesting 
point in this case is ${\bf q}_X = \pi/a (0,0,1)$, which lies on the side of the
zone.

 Consider displacements of the ions from their equilibrium positions 
of the form
\be
	\Delta b_{\kappa}^\alpha({\bf R};\eta) &=& \eta
			e^{i{\bf q}_X\cdot {\bf R}_i} 
		x_{\kappa\nu}^{\alpha}({\bf q}_X),
\en
where $\eta$ is a small real number and $\nu$ is a specific mode of interest
(X$_1$ or X$_2$).  In Fourier space, this corresponds to
\be
	\Delta b_{\kappa}^\alpha({\bf q};\eta) &=& \eta 
			\delta_{{\bf q,q}_X} 
		\sqrt{N}
		x_{\kappa\nu}^{\alpha}({\bf q}_X).
\en
 The positions of the ions, now
considered as simple numbers and not operators,
can be defined as
\be
	{\bf r}_{\kappa}({\bf R}) &=& {\bf R}+{\bf b}_\kappa
		+\Delta{\bf b}_\kappa({\bf R};\eta)+
					{\bf u}_{\kappa}({\bf R}).
\en
The dynamical matrix computed about the non-equilibrium position is
given by
\be
	D_{\kappa_1\kappa_2}^{\alpha_1\alpha_2}({\bf q};\eta)
	&=& \frac{\partial^2}
	{
	\partial u_{\kappa_1}^{\alpha_1}({\bf -q})
	\partial u_{\kappa_2}^{\alpha_2}({\bf q})}
		U[\{{\bf u}+\Delta {\bf b}\}]\Bigg|_{\bf u = 0}\\
	&=&
	D_{\kappa_1\kappa_2}^{\alpha_1\alpha_2}({\bf q})\\
	&&+\frac{\eta^2}{2}\sum_{\kappa_3,\kappa_4}\sum_{\alpha_3,\alpha_4}
		x_{\kappa_3\nu}^{\alpha_3}({\bf q}_X)
		x_{\kappa_4\nu}^{\alpha_4}({\bf q}_X)\nonumber\\
	&&
		\times\Phi_{\kappa_1\kappa_2\kappa_3\kappa_4}
		^{\alpha_1\alpha_2\alpha_3\alpha_4}
		({\bf q},{\bf -q},{\bf q}_X,{\bf q}_X)
	 +O(\eta^3);\nonumber
\en
from this last expression it is clear that the coefficients of interest
for the computation of the "loop" diagram
can be extracted from the second derivative with respect to $\eta$
of the out-of-equilibrium dynamical matrix.

The situation is slightly more complicated, however. The ionic displacements
introduced do not have the periodicity of the lattice; such periodicity
is essential in order to apply standard computational methods to extract the
dynamical matrix. The displacements are periodic with respect to a 
lattice whose cells (henceforth supercell) contain two of the original cells
stacked in the $c$ direction. These subcells of the supercells will be labeled
with $\delta = 0$ and $\delta = 1$.
 Define
\be
	{\bf A} &=& a (0,0,1);
\en
any lattice vector of the original lattice ${\bf R}$ can be expressed 
as 
\be
	{\bf R} &=& {\bf \bar R} +\delta {\bf A}
\en
where $\bf\bar R$ is some vector of the superlattice and $\delta$ can be either
0 or 1. The ionic positions with respect to this new basis are expressed
as
\be
	{\bf \bar r}_{(\kappa,\delta)}({\bf \bar R}) &=&
		{\bf \bar R}+\delta {\bf A}+
	       {\bf b}_{\kappa}+\eta e^{i\pi\delta}{\bf x}_\kappa
		+{\bf \bar u}_{(\kappa,\delta)}({\bf \bar R}),\nonumber\\
\en
where now the compound index $(\kappa,\delta)$ identifies all the ions in the
supercell with a label indicating ion $\kappa$ and subcell $\delta$.
In reciprocal space,
\be
	{\bf \bar u}_{(\kappa,\delta)}({\bf \bar R})
	&=&\sqrt{\frac{2}{N}}\sum_{\bf \bar q} e^{i\bf \bar q \cdot \bar R}
	{\bf \bar u}_{(\kappa,\delta)}({\bf \bar q})
\en
where $\bf \bar q$ is a wave vector in the 1BZ of the superlattice. The following
relationships are thus immediate:
\be
	{\bf \bar u}_{(\kappa,\delta)}({\bf \bar R}) 
			&=&	
	{\bf u}_{\kappa}({\bf R}={\bf \bar R}+\delta{\bf A}) \\
	{\bf \bar u}_{(\kappa,\delta)}({\bf \bar q}) 
			&=&	
	\frac{e^{i\delta \bf \bar q\cdot A}}{\sqrt{2}}
	\Big({\bf u}_{\kappa}({\bf \bar q })
	+e^{i\delta\pi}{\bf u}_{\kappa}({\bf \bar q+q}_X)\Big),\hspace{0.6cm}
	\nonumber\\
\en
which implies
\be
	\frac{\partial}
	{\partial \bar u_{(\kappa,\delta)}^{\alpha}({\bf \bar q})}
	=
	\frac{e^{-i\delta \bf \bar q\cdot A}}{\sqrt{2}}
	\Big(\frac{\partial}{\partial u^\alpha_{\kappa}({\bf \bar q})}
	+e^{i\delta\pi}
	\frac{\partial}{\partial u^\alpha_{\kappa}({\bf \bar q+q}_X)}\Big).\nonumber\\
\en
Above ``${\bf \bar q+q}_X$" is meant to represent 
the wave vector inside the 1BZ
of the original lattice which is obtained from ${\bf \bar q+q}_X$ by an 
appropriate reciprocal lattice translation. For what follows, it will
be useful to remember that $2\bf q_X$ is a reciprocal lattice vector of
the original system, such that all anharmonic coefficients are periodic
under ${\bf q} \rightarrow {\bf q}+2 {\bf q_X}$.

The dynamical matrix in the supercell representation is given by
\begin{widetext}
\be
	\lefteqn{
	\bar D_{(\kappa,\delta)_1,(\kappa,\delta)_2}^{\alpha_1\alpha_2}({\bf \bar q})
	= \frac{\partial^2U[\{{\bf \bar u}\}]}
	{\partial \bar u_{(\kappa,\delta)_1}^{\alpha_1}({\bf -\bar q})
	\partial \bar u_{(\kappa,\delta)_2}^{\alpha_2}({\bf \bar q})}
		\Bigg|_{\bf \bf \bar u = 0}}\\
	&=&
	\frac{e^{i(\delta_1-\delta_2)\bf \bar q\cdot A}}{2}\Bigg[
	D_{\kappa_1\kappa_2}^{\alpha_1\alpha_2}({\bf \bar q};\eta)
	+
	e^{i(\delta_1+\delta_2)\pi}
	D_{\kappa_1\kappa_2}^{\alpha_1\alpha_2}({\bf \bar q+q}_X;\eta)
	+e^{i\delta_2\pi}
	O_{\kappa_1\kappa_2}^{\alpha_1\alpha_2}({\bf \bar q};\eta)
	+
	e^{i\delta_1\pi}
	O_{\kappa_1\kappa_2}^{\alpha_1\alpha_2}({\bf \bar q+q}_X;\eta)
	\Bigg],\nonumber\\
\en
where
\be
	O_{\kappa_1\kappa_2}^{\alpha_1\alpha_2}({\bf \bar q};\eta)
		&=&
	\frac{\partial^2U[\{{\bf u}\}]}
	{\partial u_{\kappa_1}^{\alpha_1}({\bf -\bar q})
	\partial u_{\kappa_2}^{\alpha_2}({\bf \bar q+q}_X)}\Bigg|_{\bf \bf u = 0}
	= \eta \sum_{\kappa_3,\alpha_3}
		\Phi_{\kappa_1\kappa_2\kappa_3}^{\alpha_1\alpha_2\alpha_3}
		({\bf \bar q },{\bf -\bar q-q}_X,{\bf q}_X)x_{\kappa_3\nu}^{\alpha_3}({\bf q}_X)
	+O(\eta^2),
\en
and
\be
	O_{\kappa_1\kappa_2}^{\alpha_1\alpha_2}({\bf \bar q+q}_X;\eta)
		&=&
	\frac{\partial^2U[\{{\bf u}\}]}
	{\partial u_{\kappa_1}^{\alpha_1}({\bf -\bar q-q}_X)
	\partial u_{\kappa_2}^{\alpha_2}({\bf \bar q})}\Bigg|_{\bf \bf u = 0}
	= \eta \sum_{\kappa_3,\alpha_3}
		\Phi_{\kappa_1\kappa_2\kappa_3}^{\alpha_1\alpha_2\alpha_3}
		({\bf \bar q+q}_X,{\bf -\bar q},{\bf q}_X)x_{\kappa_3\nu}^{\alpha_3}({\bf q}_X)
	+O(\eta^2).
\en
\end{widetext}
From these last two terms, the anharmonic coefficients necessary for the
computation of the "bubble" diagram can be extracted. 

Thus, from the supercell dynamical matrices $\bar D(\eta)$ (the quantities
actually computed through DFPT), it is a simple matter of algebra to
extract $D(\eta)$ and $O(\eta)$. It is then useful to define partially projected
anharmonic parameters,
\be
	\Phi_{\kappa_1\kappa_2;\nu_X\nu_X}^{\alpha_1\alpha_2}({\bf q})
	&\equiv&
	\frac{\eta^2}{2}\sum_{\kappa_3,\kappa_4}\sum_{\alpha_3,\alpha_4}
		x_{\kappa_3\nu}^{\alpha_3}({\bf q}_X)
		x_{\kappa_4\nu}^{\alpha_4}({\bf q}_X)\nonumber\\
	&&
		\times\Phi_{\kappa_1\kappa_2\kappa_3\kappa_4}
		^{\alpha_1\alpha_2\alpha_3\alpha_4}
		({\bf q},{\bf -q},{\bf q}_X,{\bf q}_X)\\
	&=& \frac{\partial^2}{\partial \eta^2}
	D_{\kappa_1\kappa_2}^{\alpha_1\alpha_2}({\bf q};\eta)+O(\eta),
\en 
and
\be
	\Phi_{\kappa_1\kappa_2;\nu_X}^{\alpha_1\alpha_2}({\bf q})
	&\equiv&
	\eta\sum_{\kappa_3}\sum_{\alpha_3}
		x_{\kappa_3\nu}^{\alpha_3}({\bf q}_X)\nonumber\\
	&&
		\times\Phi_{\kappa_1\kappa_2\kappa_3}
		^{\alpha_1\alpha_2\alpha_3}
		({\bf q},{\bf -q-q_X},{\bf q}_X)\\
	&=& \frac{\partial}{\partial \eta}
	O_{\kappa_1\kappa_2}^{\alpha_1\alpha_2}({\bf q};\eta)+O(\eta).
\en 
These parameters can then straightforwardly be obtained by estimating
the $\eta$ derivatives by finite difference schemes. Furthermore,
since these coefficients are periodic in reciprocal space 
(namely periodic under $\bf q \rightarrow \bf q+ G$), they are
susceptible to Fourier interpolation, in complete analogy with
the usual procedures employed with dynamical matrices.
Once these coefficients are obtained on a dense $\bf q$ mesh, it is possible to 
compute the "loop" and "bubble" diagrams.

\bibliography{/home/bruno/work/my_documents/bibliography/general}

\begin{thebibliography}{34}%
\makeatletter
\providecommand \@ifxundefined [1]{%
 \@ifx{#1\undefined}
}%
\providecommand \@ifnum [1]{%
 \ifnum #1\expandafter \@firstoftwo
 \else \expandafter \@secondoftwo
 \fi
}%
\providecommand \@ifx [1]{%
 \ifx #1\expandafter \@firstoftwo
 \else \expandafter \@secondoftwo
 \fi
}%
\providecommand \natexlab [1]{#1}%
\providecommand \enquote  [1]{``#1''}%
\providecommand \bibnamefont  [1]{#1}%
\providecommand \bibfnamefont [1]{#1}%
\providecommand \citenamefont [1]{#1}%
\providecommand \href@noop [0]{\@secondoftwo}%
\providecommand \href [0]{\begingroup \@sanitize@url \@href}%
\providecommand \@href[1]{\@@startlink{#1}\@@href}%
\providecommand \@@href[1]{\endgroup#1\@@endlink}%
\providecommand \@sanitize@url [0]{\catcode `\\12\catcode `\$12\catcode
  `\&12\catcode `\#12\catcode `\^12\catcode `\_12\catcode `\%12\relax}%
\providecommand \@@startlink[1]{}%
\providecommand \@@endlink[0]{}%
\providecommand \url  [0]{\begingroup\@sanitize@url \@url }%
\providecommand \@url [1]{\endgroup\@href {#1}{\urlprefix }}%
\providecommand \urlprefix  [0]{URL }%
\providecommand \Eprint [0]{\href }%
\@ifxundefined \urlstyle {%
  \providecommand \doi  [0]{\begingroup \@sanitize@url \@doi}%
  \providecommand \@doi [1]{\endgroup \@@startlink {\doibase
  #1}doi:\discretionary {}{}{}#1\@@endlink }%
}{%
  \providecommand \doi  [0]{doi:\discretionary{}{}{}\begingroup
  \urlstyle{rm}\Url }%
}%
\providecommand \doibase [0]{http://dx.doi.org/}%
\providecommand \Doi [0]{\begingroup \@sanitize@url \@Doi }%
\providecommand \@Doi  [1]{\endgroup\@@startlink{\doibase#1}\@@Doi}%
\providecommand \@@Doi [1]{#1\@@endlink}%
\providecommand \selectlanguage [0]{\@gobble}%
\providecommand \bibinfo  [0]{\@secondoftwo}%
\providecommand \bibfield  [0]{\@secondoftwo}%
\providecommand \translation [1]{[#1]}%
\providecommand \BibitemOpen [0]{}%
\providecommand \bibitemStop [0]{}%
\providecommand \bibitemNoStop [0]{.\EOS\space}%
\providecommand \EOS [0]{\spacefactor3000\relax}%
\providecommand \BibitemShut  [1]{\csname bibitem#1\endcsname}%
\bibitem [{\citenamefont {Ashcroft}(1968)}]{PhysRevLett.21.1748}%
  \BibitemOpen
  \bibfield  {author} {\bibinfo {author} {\bibfnamefont {N.~W.}\ \bibnamefont
  {Ashcroft}},\ }\Doi {10.1103/PhysRevLett.21.1748} {\bibfield  {journal}
  {\bibinfo  {journal} {Phys. Rev. Lett.},\ }\textbf {\bibinfo {volume} {21}},\
  \bibinfo {pages} {1748} (\bibinfo {year} {1968})}\BibitemShut {NoStop}%
\bibitem [{\citenamefont {Cudazzo}\ \emph {et~al.}(2008)\citenamefont
  {Cudazzo}, \citenamefont {Profeta}, \citenamefont {Sanna}, \citenamefont
  {Floris} \emph {et~al.}}]{cudazzo:257001}%
  \BibitemOpen
  \bibfield  {author} {\bibinfo {author} {\bibfnamefont {P.}~\bibnamefont
  {Cudazzo}}, \bibinfo {author} {\bibfnamefont {G.}~\bibnamefont {Profeta}},
  \bibinfo {author} {\bibfnamefont {A.}~\bibnamefont {Sanna}}, \bibinfo
  {author} {\bibfnamefont {A.}~\bibnamefont {Floris}},  \emph {et~al.},\
  }\href@noop {} {\bibfield  {journal} {\bibinfo  {journal} {Phys. Rev.
  Lett.},\ }\textbf {\bibinfo {volume} {100}},\ \bibinfo {pages} {257001}
  (\bibinfo {year} {2008})}\BibitemShut {NoStop}%
\bibitem [{\citenamefont {Babaev}\ \emph {et~al.}(2005)\citenamefont {Babaev},
  \citenamefont {Sudbo},\ and\ \citenamefont {Ashcroft}}]{babaev:105301}%
  \BibitemOpen
  \bibfield  {author} {\bibinfo {author} {\bibfnamefont {E.}~\bibnamefont
  {Babaev}}, \bibinfo {author} {\bibfnamefont {A.}~\bibnamefont {Sudbo}}, \
  and\ \bibinfo {author} {\bibfnamefont {N.~W.}\ \bibnamefont {Ashcroft}},\
  }\href@noop {} {\bibfield  {journal} {\bibinfo  {journal} {Phys. Rev.
  Lett.},\ }\textbf {\bibinfo {volume} {95}},\ \bibinfo {eid} {105301}
  (\bibinfo {year} {2005})}\BibitemShut {NoStop}%
\bibitem [{\citenamefont {Babaev}\ and\ \citenamefont
  {Ashcroft}(2007)}]{babaev_nature}%
  \BibitemOpen
  \bibfield  {author} {\bibinfo {author} {\bibfnamefont {E.}~\bibnamefont
  {Babaev}}\ and\ \bibinfo {author} {\bibfnamefont {N.~W.}\ \bibnamefont
  {Ashcroft}},\ }\href@noop {} {\bibfield  {journal} {\bibinfo  {journal} {Nat.
  Phys.},\ }\textbf {\bibinfo {volume} {3}},\ \bibinfo {pages} {530} (\bibinfo
  {year} {2007})}\BibitemShut {NoStop}%
\bibitem [{\citenamefont {Nagao}\ \emph {et~al.}(2003)\citenamefont {Nagao},
  \citenamefont {Bonev}, \citenamefont {Bergara},\ and\ \citenamefont
  {Ashcroft}}]{PhysRevLett.90.035501}%
  \BibitemOpen
  \bibfield  {author} {\bibinfo {author} {\bibfnamefont {K.}~\bibnamefont
  {Nagao}}, \bibinfo {author} {\bibfnamefont {S.~A.}\ \bibnamefont {Bonev}},
  \bibinfo {author} {\bibfnamefont {A.}~\bibnamefont {Bergara}}, \ and\
  \bibinfo {author} {\bibfnamefont {N.~W.}\ \bibnamefont {Ashcroft}},\ }\Doi
  {10.1103/PhysRevLett.90.035501} {\bibfield  {journal} {\bibinfo  {journal}
  {Phys. Rev. Lett.},\ }\textbf {\bibinfo {volume} {90}},\ \bibinfo {pages}
  {035501} (\bibinfo {year} {2003})}\BibitemShut {NoStop}%
\bibitem [{\citenamefont {Loubeyre}\ \emph {et~al.}(2002)\citenamefont
  {Loubeyre}, \citenamefont {Occelli},\ and\ \citenamefont
  {LeToullec}}]{Loubeyre_Hydrogen_nature}%
  \BibitemOpen
  \bibfield  {author} {\bibinfo {author} {\bibfnamefont {P.}~\bibnamefont
  {Loubeyre}}, \bibinfo {author} {\bibfnamefont {F.}~\bibnamefont {Occelli}}, \
  and\ \bibinfo {author} {\bibfnamefont {R.}~\bibnamefont {LeToullec}},\
  }\href@noop {} {\bibfield  {journal} {\bibinfo  {journal} {Nature},\ }\textbf
  {\bibinfo {volume} {416}},\ \bibinfo {pages} {613} (\bibinfo {year}
  {2002})}\BibitemShut {NoStop}%
\bibitem [{\citenamefont {Ashcroft}(2004)}]{PhysRevLett.92.187002}%
  \BibitemOpen
  \bibfield  {author} {\bibinfo {author} {\bibfnamefont {N.~W.}\ \bibnamefont
  {Ashcroft}},\ }\Doi {10.1103/PhysRevLett.92.187002} {\bibfield  {journal}
  {\bibinfo  {journal} {Phys. Rev. Lett.},\ }\textbf {\bibinfo {volume} {92}},\
  \bibinfo {pages} {187002} (\bibinfo {year} {2004})}\BibitemShut {NoStop}%
\bibitem [{\citenamefont {Feng}\ \emph {et~al.}(2006)\citenamefont {Feng},
  \citenamefont {Grochala}, \citenamefont {Jaro\'{n}}, \citenamefont {Hoffmann}
  \emph {et~al.}}]{feng:017006}%
  \BibitemOpen
  \bibfield  {author} {\bibinfo {author} {\bibfnamefont {J.}~\bibnamefont
  {Feng}}, \bibinfo {author} {\bibfnamefont {W.}~\bibnamefont {Grochala}},
  \bibinfo {author} {\bibfnamefont {T.}~\bibnamefont {Jaro\'{n}}}, \bibinfo
  {author} {\bibfnamefont {R.}~\bibnamefont {Hoffmann}},  \emph {et~al.},\
  }\Doi {10.1103/PhysRevLett.96.017006} {\bibfield  {journal} {\bibinfo
  {journal} {Phys. Rev. Lett.},\ }\textbf {\bibinfo {volume} {96}},\ \bibinfo
  {eid} {017006} (\bibinfo {year} {2006})}\BibitemShut {NoStop}%
\bibitem [{\citenamefont {Gao}\ \emph {et~al.}(2010)\citenamefont {Gao},
  \citenamefont {Oganov}, \citenamefont {Li}, \citenamefont {Li} \emph
  {et~al.}}]{stannane}%
  \BibitemOpen
  \bibfield  {author} {\bibinfo {author} {\bibfnamefont {G.}~\bibnamefont
  {Gao}}, \bibinfo {author} {\bibfnamefont {A.~R.}\ \bibnamefont {Oganov}},
  \bibinfo {author} {\bibfnamefont {P.}~\bibnamefont {Li}}, \bibinfo {author}
  {\bibfnamefont {Z.}~\bibnamefont {Li}},  \emph {et~al.},\ }\href@noop {}
  {\bibfield  {journal} {\bibinfo  {journal} {PNAS},\ }\textbf {\bibinfo
  {volume} {26}},\ \bibinfo {pages} {1317} (\bibinfo {year}
  {2010})}\BibitemShut {NoStop}%
\bibitem [{\citenamefont {Martinez-Canales}\ \emph {et~al.}(2009)\citenamefont
  {Martinez-Canales}, \citenamefont {Oganov}, \citenamefont {Ma}, \citenamefont
  {Yan} \emph {et~al.}}]{PhysRevLett.102.087005}%
  \BibitemOpen
  \bibfield  {author} {\bibinfo {author} {\bibfnamefont {M.}~\bibnamefont
  {Martinez-Canales}}, \bibinfo {author} {\bibfnamefont {A.~R.}\ \bibnamefont
  {Oganov}}, \bibinfo {author} {\bibfnamefont {Y.}~\bibnamefont {Ma}}, \bibinfo
  {author} {\bibfnamefont {Y.}~\bibnamefont {Yan}},  \emph {et~al.},\ }\Doi
  {10.1103/PhysRevLett.102.087005} {\bibfield  {journal} {\bibinfo  {journal}
  {Phys. Rev. Lett.},\ }\textbf {\bibinfo {volume} {102}},\ \bibinfo {pages}
  {087005} (\bibinfo {year} {2009})}\BibitemShut {NoStop}%
\bibitem [{\citenamefont {Gao}\ \emph {et~al.}(2008)\citenamefont {Gao},
  \citenamefont {Oganov}, \citenamefont {Bergara}, \citenamefont
  {Martinez-Canales} \emph {et~al.}}]{PhysRevLett.101.107002}%
  \BibitemOpen
  \bibfield  {author} {\bibinfo {author} {\bibfnamefont {G.}~\bibnamefont
  {Gao}}, \bibinfo {author} {\bibfnamefont {A.~R.}\ \bibnamefont {Oganov}},
  \bibinfo {author} {\bibfnamefont {A.}~\bibnamefont {Bergara}}, \bibinfo
  {author} {\bibfnamefont {M.}~\bibnamefont {Martinez-Canales}},  \emph
  {et~al.},\ }\Doi {10.1103/PhysRevLett.101.107002} {\bibfield  {journal}
  {\bibinfo  {journal} {Phys. Rev. Lett.},\ }\textbf {\bibinfo {volume}
  {101}},\ \bibinfo {pages} {107002} (\bibinfo {year} {2008})}\BibitemShut
  {NoStop}%
\bibitem [{\citenamefont {Martinez-Canales}\ \emph {et~al.}(2006)\citenamefont
  {Martinez-Canales}, \citenamefont {Bergara}, \citenamefont {Feng},\ and\
  \citenamefont {Grochala}}]{MartinezCanales20062095}%
  \BibitemOpen
  \bibfield  {author} {\bibinfo {author} {\bibfnamefont {M.}~\bibnamefont
  {Martinez-Canales}}, \bibinfo {author} {\bibfnamefont {A.}~\bibnamefont
  {Bergara}}, \bibinfo {author} {\bibfnamefont {J.}~\bibnamefont {Feng}}, \
  and\ \bibinfo {author} {\bibfnamefont {W.}~\bibnamefont {Grochala}},\
  }\href@noop {} {\bibfield  {journal} {\bibinfo  {journal} {J. Phys. Chem.
  Solids},\ }\textbf {\bibinfo {volume} {67}},\ \bibinfo {pages} {2095 }
  (\bibinfo {year} {2006})}\BibitemShut {NoStop}%
\bibitem [{\citenamefont {Degtyareva}\ \emph {et~al.}(2007)\citenamefont
  {Degtyareva}, \citenamefont {Canales}, \citenamefont {Bergara}, \citenamefont
  {Chen} \emph {et~al.}}]{PhysRevB.76.064123}%
  \BibitemOpen
  \bibfield  {author} {\bibinfo {author} {\bibfnamefont {O.}~\bibnamefont
  {Degtyareva}}, \bibinfo {author} {\bibfnamefont {M.~M.}\ \bibnamefont
  {Canales}}, \bibinfo {author} {\bibfnamefont {A.}~\bibnamefont {Bergara}},
  \bibinfo {author} {\bibfnamefont {X.-J.}\ \bibnamefont {Chen}},  \emph
  {et~al.},\ }\Doi {10.1103/PhysRevB.76.064123} {\bibfield  {journal} {\bibinfo
   {journal} {Phys. Rev. B},\ }\textbf {\bibinfo {volume} {76}},\ \bibinfo
  {pages} {064123} (\bibinfo {year} {2007})}\BibitemShut {NoStop}%
\bibitem [{\citenamefont {Eremets}\ \emph {et~al.}(2008)\citenamefont
  {Eremets}, \citenamefont {Trojan}, \citenamefont {Medvedev}, \citenamefont
  {Tse} \emph {et~al.}}]{Eremets_sih4}%
  \BibitemOpen
  \bibfield  {author} {\bibinfo {author} {\bibfnamefont {M.~I.}\ \bibnamefont
  {Eremets}}, \bibinfo {author} {\bibfnamefont {I.~A.}\ \bibnamefont {Trojan}},
  \bibinfo {author} {\bibfnamefont {S.~A.}\ \bibnamefont {Medvedev}}, \bibinfo
  {author} {\bibfnamefont {J.~S.}\ \bibnamefont {Tse}},  \emph {et~al.},\
  }\href@noop {} {\bibfield  {journal} {\bibinfo  {journal} {Science},\
  }\textbf {\bibinfo {volume} {319}},\ \bibinfo {pages} {1506} (\bibinfo {year}
  {2008})}\BibitemShut {NoStop}%
\bibitem [{\citenamefont {Chen}\ \emph {et~al.}(2008)\citenamefont {Chen},
  \citenamefont {Struzhkin}, \citenamefont {Song}, \citenamefont {Goncharov}
  \emph {et~al.}}]{metalization_sih4_pnas}%
  \BibitemOpen
  \bibfield  {author} {\bibinfo {author} {\bibfnamefont {X.-J.}\ \bibnamefont
  {Chen}}, \bibinfo {author} {\bibfnamefont {V.~V.}\ \bibnamefont {Struzhkin}},
  \bibinfo {author} {\bibfnamefont {Y.}~\bibnamefont {Song}}, \bibinfo {author}
  {\bibfnamefont {A.~F.}\ \bibnamefont {Goncharov}},  \emph {et~al.},\
  }\href@noop {} {\bibfield  {journal} {\bibinfo  {journal} {PNAS},\ }\textbf
  {\bibinfo {volume} {105}},\ \bibinfo {pages} {20} (\bibinfo {year}
  {2008})}\BibitemShut {NoStop}%
\bibitem [{\citenamefont {Kim}\ \emph {et~al.}(2010)\citenamefont {Kim},
  \citenamefont {Scheicher}, \citenamefont {kwang Mao}, \citenamefont {Kang}
  \emph {et~al.}}]{Ahuja_trends}%
  \BibitemOpen
  \bibfield  {author} {\bibinfo {author} {\bibfnamefont {D.~Y.}\ \bibnamefont
  {Kim}}, \bibinfo {author} {\bibfnamefont {R.~H.}\ \bibnamefont {Scheicher}},
  \bibinfo {author} {\bibfnamefont {H.}~\bibnamefont {kwang Mao}}, \bibinfo
  {author} {\bibfnamefont {T.~W.}\ \bibnamefont {Kang}},  \emph {et~al.},\
  }\href@noop {} {\bibfield  {journal} {\bibinfo  {journal} {PNAS},\ }\textbf
  {\bibinfo {volume} {107}},\ \bibinfo {pages} {2793 } (\bibinfo {year}
  {2010})}\BibitemShut {NoStop}%
\bibitem [{\citenamefont {Pickard}\ and\ \citenamefont
  {Needs}(2007)}]{pickard:144114}%
  \BibitemOpen
  \bibfield  {author} {\bibinfo {author} {\bibfnamefont {C.~J.}\ \bibnamefont
  {Pickard}}\ and\ \bibinfo {author} {\bibfnamefont {R.~J.}\ \bibnamefont
  {Needs}},\ }\href@noop {} {\bibfield  {journal} {\bibinfo  {journal} {Phys.
  Rev. B},\ }\textbf {\bibinfo {volume} {76}},\ \bibinfo {eid} {144114}
  (\bibinfo {year} {2007})}\BibitemShut {NoStop}%
\bibitem [{\citenamefont {Kim}\ \emph {et~al.}(2008)\citenamefont {Kim},
  \citenamefont {Scheicher},\ and\ \citenamefont {Ahuja}}]{kim:100102}%
  \BibitemOpen
  \bibfield  {author} {\bibinfo {author} {\bibfnamefont {D.~Y.}\ \bibnamefont
  {Kim}}, \bibinfo {author} {\bibfnamefont {R.~H.}\ \bibnamefont {Scheicher}},
  \ and\ \bibinfo {author} {\bibfnamefont {R.}~\bibnamefont {Ahuja}},\ }\Doi
  {10.1103/PhysRevB.78.100102} {\bibfield  {journal} {\bibinfo  {journal}
  {Phys. Rev. B},\ }\textbf {\bibinfo {volume} {78}},\ \bibinfo {eid} {100102}
  (\bibinfo {year} {2008})}\BibitemShut {NoStop}%
\bibitem [{\citenamefont {Goncharenko}\ \emph {et~al.}(2008)\citenamefont
  {Goncharenko}, \citenamefont {Eremets}, \citenamefont {Hanfland},
  \citenamefont {Tse} \emph {et~al.}}]{goncharenko:045504}%
  \BibitemOpen
  \bibfield  {author} {\bibinfo {author} {\bibfnamefont {I.}~\bibnamefont
  {Goncharenko}}, \bibinfo {author} {\bibfnamefont {M.~I.}\ \bibnamefont
  {Eremets}}, \bibinfo {author} {\bibfnamefont {M.}~\bibnamefont {Hanfland}},
  \bibinfo {author} {\bibfnamefont {J.~S.}\ \bibnamefont {Tse}},  \emph
  {et~al.},\ }\href@noop {} {\bibfield  {journal} {\bibinfo  {journal} {Phys.
  Rev. Lett.},\ }\textbf {\bibinfo {volume} {100}},\ \bibinfo {eid} {045504}
  (\bibinfo {year} {2008})}\BibitemShut {NoStop}%
\bibitem [{\citenamefont {Allen}\ and\ \citenamefont
  {Mitrovic}(1982)}]{AllenSSP}%
  \BibitemOpen
  \bibfield  {author} {\bibinfo {author} {\bibfnamefont {P.~B.}\ \bibnamefont
  {Allen}}\ and\ \bibinfo {author} {\bibfnamefont {B.}~\bibnamefont
  {Mitrovic}},\ }\href@noop {} {\bibfield  {journal} {\bibinfo  {journal}
  {Solid State Physics},\ }\textbf {\bibinfo {volume} {37}},\ \bibinfo {pages}
  {1} (\bibinfo {year} {1982})}\BibitemShut {NoStop}%
\bibitem [{\citenamefont {McMillan}(1968)}]{PhysRev.167.331}%
  \BibitemOpen
  \bibfield  {author} {\bibinfo {author} {\bibfnamefont {W.~L.}\ \bibnamefont
  {McMillan}},\ }\href@noop {} {\bibfield  {journal} {\bibinfo  {journal}
  {Phys. Rev.},\ }\textbf {\bibinfo {volume} {167}},\ \bibinfo {pages} {331}
  (\bibinfo {year} {1968})}\BibitemShut {NoStop}%
\bibitem [{\citenamefont {Allen}\ and\ \citenamefont
  {Dynes}(1975)}]{PhysRevB.12.905}%
  \BibitemOpen
  \bibfield  {author} {\bibinfo {author} {\bibfnamefont {P.~B.}\ \bibnamefont
  {Allen}}\ and\ \bibinfo {author} {\bibfnamefont {R.~C.}\ \bibnamefont
  {Dynes}},\ }\href@noop {} {\bibfield  {journal} {\bibinfo  {journal} {Phys.
  Rev. B},\ }\textbf {\bibinfo {volume} {12}},\ \bibinfo {pages} {905}
  (\bibinfo {year} {1975})}\BibitemShut {NoStop}%
\bibitem [{\citenamefont {Allen}(1972)}]{PhysRevB.6.2577}%
  \BibitemOpen
  \bibfield  {author} {\bibinfo {author} {\bibfnamefont {P.~B.}\ \bibnamefont
  {Allen}},\ }\href@noop {} {\bibfield  {journal} {\bibinfo  {journal} {Phys.
  Rev. B},\ }\textbf {\bibinfo {volume} {6}},\ \bibinfo {pages} {2577}
  (\bibinfo {year} {1972})}\BibitemShut {NoStop}%
\bibitem [{\citenamefont {Mahan}(2000)}]{Mahan}%
  \BibitemOpen
  \bibfield  {author} {\bibinfo {author} {\bibfnamefont {G.~D.}\ \bibnamefont
  {Mahan}},\ }\href@noop {} {\emph {\bibinfo {title} {Many-Particle Physics}}}\
  (\bibinfo  {publisher} {Klumer Academic},\ \bibinfo {year}
  {2000})\BibitemShut {NoStop}%
\bibitem [{\citenamefont {Maradudin}\ and\ \citenamefont
  {Fein}(1962)}]{PhysRev.128.2589}%
  \BibitemOpen
  \bibfield  {author} {\bibinfo {author} {\bibfnamefont {A.~A.}\ \bibnamefont
  {Maradudin}}\ and\ \bibinfo {author} {\bibfnamefont {A.~E.}\ \bibnamefont
  {Fein}},\ }\Doi {10.1103/PhysRev.128.2589} {\bibfield  {journal} {\bibinfo
  {journal} {Phys. Rev.},\ }\textbf {\bibinfo {volume} {128}},\ \bibinfo
  {pages} {2589} (\bibinfo {year} {1962})}\BibitemShut {NoStop}%
\bibitem [{\citenamefont {Giannozzi}\ \emph {et~al.}(2009)\citenamefont
  {Giannozzi}, \citenamefont {Baroni}, \citenamefont {Bonini}, \citenamefont
  {Calandra} \emph {et~al.}}]{QE-2009}%
  \BibitemOpen
  \bibfield  {author} {\bibinfo {author} {\bibfnamefont {P.}~\bibnamefont
  {Giannozzi}}, \bibinfo {author} {\bibfnamefont {S.}~\bibnamefont {Baroni}},
  \bibinfo {author} {\bibfnamefont {N.}~\bibnamefont {Bonini}}, \bibinfo
  {author} {\bibfnamefont {M.}~\bibnamefont {Calandra}},  \emph {et~al.},\
  }\href@noop {} {\bibfield  {journal} {\bibinfo  {journal} {J. Phys. Condens.
  Matter},\ }\textbf {\bibinfo {volume} {21}},\ \bibinfo {pages} {395502}
  (\bibinfo {year} {2009})}\BibitemShut {NoStop}%
\bibitem [{\citenamefont {Perdew}\ \emph {et~al.}(1997)\citenamefont {Perdew},
  \citenamefont {Burke},\ and\ \citenamefont
  {Ernzerhof}}]{PhysRevLett.78.1396}%
  \BibitemOpen
  \bibfield  {author} {\bibinfo {author} {\bibfnamefont {J.~P.}\ \bibnamefont
  {Perdew}}, \bibinfo {author} {\bibfnamefont {K.}~\bibnamefont {Burke}}, \
  and\ \bibinfo {author} {\bibfnamefont {M.}~\bibnamefont {Ernzerhof}},\ }\Doi
  {10.1103/PhysRevLett.78.1396} {\bibfield  {journal} {\bibinfo  {journal}
  {Phys. Rev. Lett.},\ }\textbf {\bibinfo {volume} {78}},\ \bibinfo {pages}
  {1396} (\bibinfo {year} {1997})}\BibitemShut {NoStop}%
\bibitem [{\citenamefont {Perdew}\ \emph {et~al.}(1996)\citenamefont {Perdew},
  \citenamefont {Burke},\ and\ \citenamefont
  {Ernzerhof}}]{PhysRevLett.77.3865}%
  \BibitemOpen
  \bibfield  {author} {\bibinfo {author} {\bibfnamefont {J.~P.}\ \bibnamefont
  {Perdew}}, \bibinfo {author} {\bibfnamefont {K.}~\bibnamefont {Burke}}, \
  and\ \bibinfo {author} {\bibfnamefont {M.}~\bibnamefont {Ernzerhof}},\ }\Doi
  {10.1103/PhysRevLett.77.3865} {\bibfield  {journal} {\bibinfo  {journal}
  {Phys. Rev. Lett.},\ }\textbf {\bibinfo {volume} {77}},\ \bibinfo {pages}
  {3865} (\bibinfo {year} {1996})}\BibitemShut {NoStop}%
\bibitem [{\citenamefont {Vanderbilt}(1990)}]{PhysRevB.41.7892}%
  \BibitemOpen
  \bibfield  {author} {\bibinfo {author} {\bibfnamefont {D.}~\bibnamefont
  {Vanderbilt}},\ }\href@noop {} {\bibfield  {journal} {\bibinfo  {journal}
  {Phys. Rev. B},\ }\textbf {\bibinfo {volume} {41}},\ \bibinfo {pages} {7892}
  (\bibinfo {year} {1990})}\BibitemShut {NoStop}%
\bibitem [{\citenamefont {Baroni}\ \emph {et~al.}(2001)\citenamefont {Baroni},
  \citenamefont {de~Gironcoli}, \citenamefont {Dal~Corso},\ and\ \citenamefont
  {Giannozzi}}]{dfpt}%
  \BibitemOpen
  \bibfield  {author} {\bibinfo {author} {\bibfnamefont {S.}~\bibnamefont
  {Baroni}}, \bibinfo {author} {\bibfnamefont {S.}~\bibnamefont
  {de~Gironcoli}}, \bibinfo {author} {\bibfnamefont {A.}~\bibnamefont
  {Dal~Corso}}, \ and\ \bibinfo {author} {\bibfnamefont {P.}~\bibnamefont
  {Giannozzi}},\ }\href@noop {} {\bibfield  {journal} {\bibinfo  {journal}
  {Rev. Mod. Phys.},\ }\textbf {\bibinfo {volume} {73}},\ \bibinfo {pages}
  {515} (\bibinfo {year} {2001})}\BibitemShut {NoStop}%
\bibitem [{Note1()}]{Note1}%
  \BibitemOpen
  \bibinfo {note} {To get a sense of scale, note that, for the ion most
  displaced, $|{\protect \bf x}_{\kappa \nu _x}|\simeq 0.15\protect \tmspace
  +\thinmuskip {.1667em}a_0 \simeq 0.03\protect \tmspace +\thinmuskip
  {.1667em}a$ for X$_1$ and $|{\protect \bf x}_{\kappa \nu _x}|\simeq
  0.14\protect \tmspace +\thinmuskip {.1667em}a_0 \simeq 0.02\protect \tmspace
  +\thinmuskip {.1667em}a$ for X$_2$.}\BibitemShut {Stop}%
\bibitem [{\citenamefont {Yildirim}\ \emph {et~al.}(2001)\citenamefont
  {Yildirim}, \citenamefont {G\"ulseren}, \citenamefont {Lynn},\ and\
  \citenamefont {\textit{et al.}}}]{PhysRevLett.87.037001}%
  \BibitemOpen
  \bibfield  {author} {\bibinfo {author} {\bibfnamefont {T.}~\bibnamefont
  {Yildirim}}, \bibinfo {author} {\bibfnamefont {O.}~\bibnamefont
  {G\"ulseren}}, \bibinfo {author} {\bibfnamefont {J.~W.}\ \bibnamefont
  {Lynn}}, \ and\ \bibinfo {author} {\bibfnamefont {B.}~\bibnamefont
  {\textit{et al.}}},\ }\Doi {10.1103/PhysRevLett.87.037001} {\bibfield
  {journal} {\bibinfo  {journal} {Phys. Rev. Lett.},\ }\textbf {\bibinfo
  {volume} {87}},\ \bibinfo {pages} {037001} (\bibinfo {year}
  {2001})}\BibitemShut {NoStop}%
\bibitem [{\citenamefont {Lazzeri}\ \emph {et~al.}(2003)\citenamefont
  {Lazzeri}, \citenamefont {Calandra},\ and\ \citenamefont
  {Mauri}}]{PhysRevB.68.220509}%
  \BibitemOpen
  \bibfield  {author} {\bibinfo {author} {\bibfnamefont {M.}~\bibnamefont
  {Lazzeri}}, \bibinfo {author} {\bibfnamefont {M.}~\bibnamefont {Calandra}}, \
  and\ \bibinfo {author} {\bibfnamefont {F.}~\bibnamefont {Mauri}},\ }\Doi
  {10.1103/PhysRevB.68.220509} {\bibfield  {journal} {\bibinfo  {journal}
  {Phys. Rev. B},\ }\textbf {\bibinfo {volume} {68}},\ \bibinfo {pages}
  {220509} (\bibinfo {year} {2003})}\BibitemShut {NoStop}%
\bibitem [{\citenamefont {d'Astuto}\ \emph {et~al.}(2007)\citenamefont
  {d'Astuto}, \citenamefont {Calandra}, \citenamefont {Reich}, \citenamefont
  {Shukla} \emph {et~al.}}]{d'astuto:174508}%
  \BibitemOpen
  \bibfield  {author} {\bibinfo {author} {\bibfnamefont {M.}~\bibnamefont
  {d'Astuto}}, \bibinfo {author} {\bibfnamefont {M.}~\bibnamefont {Calandra}},
  \bibinfo {author} {\bibfnamefont {S.}~\bibnamefont {Reich}}, \bibinfo
  {author} {\bibfnamefont {A.}~\bibnamefont {Shukla}},  \emph {et~al.},\ }\Doi
  {10.1103/PhysRevB.75.174508} {\bibfield  {journal} {\bibinfo  {journal}
  {Phys. Rev. B},\ }\textbf {\bibinfo {volume} {75}},\ \bibinfo {eid} {174508}
  (\bibinfo {year} {2007})}\BibitemShut {NoStop}%
\end{thebibliography}%

\end{document}